\documentclass[pra, 10pt, twocolumn,letter, floatfix]{revtex4}
\usepackage{amsmath}
\usepackage{amsfonts}
\usepackage{amssymb}
\usepackage{graphicx}
\usepackage{rotating}
\allowdisplaybreaks
\usepackage{setspace}
\usepackage{bm}

\begin{document}

\title{The Jastrow antisymmetric geminal power in Hilbert space:  theory, benchmarking, and application to a novel transition state}

\author{Eric Neuscamman\footnote[1]{Electronic mail: eric.neuscamman@gmail.com}}
\affiliation{Department of Chemistry, University of California, Berkeley, California 94720, USA}

\date{\today}

\begin{abstract}
The Jastrow-modified antisymmetric geminal power (JAGP) ansatz in Hilbert space successfully overcomes
two key failings of other pairing theories, namely a lack of inter-pair correlations and a lack of
multiple resonance structures, while maintaining a polynomially scaling cost, variational energies,
and size consistency.
Here we present efficient quantum Monte Carlo algorithms that evaluate and optimize the JAGP energy
for a cost that scales as the fifth power of the system size.
We demonstrate the JAGP's ability to describe both static and dynamic correlation by applying it
to bond stretching in H$_2$O, C$_2$, and N$_2$ as well as to a novel, multi-reference transition
state of ethene.
JAGP's accuracy in these systems outperforms even the most sophisticated single-reference methods
and approaches that of exponentially-scaling active space methods.
\end{abstract}

\maketitle

\section{Introduction}
\label{sec:introduction}

%We worked on stuff.  We should probably cite \cite{Sorella:2004:agp_sr}.

In predicting the effects of correlations between the positions of a molecule's electrons,
one must go beyond the independent particle approximation embodied by the Slater determinant
ansatz.
%, which is an antisymmetrized product of one-electron functions that encodes only those
%correlations required by fermionic antisymmetry.
In most quantum chemical methods used today, correlations are accounted for in an
essentially perturbative manner:  Kohn-Sham density functional theory \cite{Parr-Yang},
M{\o}ller-Plesset perturbation theory \cite{Szabo-Ostland}, and single-reference coupled cluster theory
\cite{BARTLETT:2007:cc_review} all rely on the fundamental assumption that
electron correlation does not greatly alter the wave function structure beyond what can be
encoded in a single Slater determinant.
This assumption can be used to define two classes of electron correlation.
Weak, or dynamic, correlations are those that can be incorporated accurately under this essentially
perturbative framework.
Strong, or static, correlations, on the other hand, are those that cannot be accounted for by
perturbing around a single Slater determinant.
While this division is certainly not the only useful way to distinguish different classes of
electron correlation, it will serve as the definition of weak and strong for the purposes of this
manuscript.

Traditionally, the strong electron correlations encountered in excited states, bond dissociations,
and transition metal chemistry have been treated using active space theories
\cite{Helgaker_book}.
Instead of dividing molecular orbitals into occupied and unoccupied categories based on the
presence of a large gap in the orbital energy spectrum, which is typically lacking in systems with
strong electron correlations, these methods divide orbitals into three categories:  low energy
orbitals that can be safely assumed to be doubly occupied, high energy orbitals that can be
assumed empty, and a set of orbitals near the Fermi energy that admit multiple competitive
electronic configurations.
While this active space is of course much smaller than the full Hilbert space, it still
grows combinatorially with system size, causing methods that rely on it to scale exponentially.
%The Hamiltonian is then diagonalized in the space of all possible electron configurations
%in this active space, whose number will in general grow exponentially in system size.
%This diagonalization produces a zeroth order reference wave function, from which similar
%techniques to those used with Slater determinants are used to incorperate any remaining
%electron correlations \cite{Knowles:1988:mrci,Werner:1988:mrci,Werner:1996:caspt2}.
%Although recent advances such as the density matrix renormalization group 
%\cite{Chan:2010:mr_dmrg} have greatly expanded the size of the active space that can be
%treated, their costs still scale exponentially in system size except in special cases
%(e.g. one dimensional systems).

Here we pursue an alternative approach to electron correlation that does not rely on either the
independent particle approximation or the use of an active space.
Instead, the conceptual starting point is a molecule whose electrons are organized into a
valence bonding pattern constructed from a product of two-electron units (bonds, lone pairs, etc.).
This picture is encoded by the perfect pairing (PP) ansatz \cite{POPLE:1953:agp} and its generalizations,
in which the wave function is constructed explicitly as an antisymmetric product of two-electron functions
(geminals).
This approach can be seen as a direct generalization of the Slater determinant, which is product of
one-electron functions (orbitals), that seeks to treat electron correlation in a pairwise manner.

There are, however, two major problems with the perfect pairing approach.
First, in order to be computationally feasible, additional approximations must be made in which
the geminals are constrained to be strongly \cite{POPLE:1953:agp} or at least partially
\cite{Bultinck:2013:nonorth_gems} orthogonal to each other,
which reduces the ansatz's ability to capture inter-pair electron correlations
\cite{Kutzelnigg:1999:cumulants}.
Second, the organization of electrons into a single pairing structure makes it difficult to describe
wave functions that resonate between multiple valence bonding patterns.
This shortcoming causes PP to exhibit sharp cusps in its potential energy surfaces
when the energies of solutions corresponding to different pairing motifs cross \cite{Parkhill:2009:perfect_quads}.
Many attempts to repair these problems have been made, including theories based on
configuration interaction \cite{Goddard:1975:gvbci,Surjan:2012:apsg},
coupled cluster \cite{Parkhill:2009:perfect_quads,Parkhill:2010:perfect_hexs,Small:2009:ccvb},
perturbation theory \cite{Surjan:2012:apsg},
and Hopf algebra \cite{Cassam:2003:hopf_alg,Cassam:2006:hopf_alg}.
However, these approaches all violate one or more of PP's important formal properties:
size consistency (additive energies of separated subsystems), variational energies, and polynomial cost.

This paper will show that by using the Hilbert space version \cite{Neuscamman:2012:sc_jagp} of the
Jastrow-modified antisymmetric geminal power (JAGP) ansatz
\cite{Sorella:2003:agp_sr,Sorella:2004:agp_sr,Sorella:2007:jagp_vdw,Sorella:2009:jagp_molec},
both inter-pair correlations and multiple resonance structures can be described without
sacrificing the variational principle, size consistency, or polynomial cost.
On its own, the geminal power can create a superposition of multiple resonance structures using non-orthogonal
local geminals.
However, in addition to the resonance structures we want, this superposition will typically also contain
unphysical resonance structures in which too many electrons are present on one bond or atom.
By incorporating the Hilbert space Jastrow factor, which can delete terms with undesirable electron
concentrations \cite{Neuscamman:2012:sc_jagp}, we prune away these unphysical resonance structures
to produce the desired superposition of physically relevant valence bond structures, each of which is
a product of several non-orthogonal local geminals.
We will show that this ansatz captures the most important electron correlations in a variety of strongly
correlated molecular examples, and that it can be optimized for a cost that scales polynomially as the fifth
power of the system size.

This paper is organized as follows.
We begin with an overview of the JAGP ansatz and the variational Monte Carlo (VMC) framework
in which our methods are based (Section \ref{sec:basics}).
We then discuss the slightly non-standard way in which we organize the terms of the ab initio
Hamiltonian (Section \ref{sec:hamiltonian}) before presenting the polynomial cost algorithms for
evaluating (Sections \ref{sec:agp_energy} and \ref{sec:jagp_energy})
and optimizing (Sections \ref{sec:derivatives} and \ref{sec:optimization}) the JAGP energy.
We then present results showing the convergence behavior of the optimization
(Section \ref{sec:convergence}), performance in a pedegogical example involving
H$_4$ (Section \ref{sec:hydrogen}), and the accuracy of the method for singlet-state multi-bond
stretching (Sections \ref{sec:h2o} and \ref{sec:n2}), triplet-state multi-bond stretching
(Section \ref{sec:c2}), and barrier height evaluation for a novel ethene hydrogenation mechanism
(Section \ref{sec:ethene}).
We conclude (Section \ref{sec:conclusions}) with a summary of our findings and comments on
the future development of the JAGP.

\begin{table}[t]
\centering
\caption{Notation}
\label{tab:defs}
\begin{tabular}{  c  l  }
\hline\hline
%\hspace{0mm} Variable \hspace{0mm} &
%\hspace{0mm} Definition \hspace{0mm} \\
%\hline
%$|\Psi\rangle$ &  $\qquad$ the wave function \\ 
$|\bm{\mathrm{n}}\rangle$ &  $\qquad$ an occupation number vector \\ 
$n_o$ &  $\qquad$ number of $\alpha$ orbitals occupied in $|\bm{\mathrm{n}}\rangle$ \\ 
$n_u$ &  $\qquad$ number of $\alpha$ orbitals unoccupied in $|\bm{\mathrm{n}}\rangle$ \\ 
$n_s$ &  $\qquad$ number of Monte Carlo samples \\ 
$p,q,r,s$ &  $\qquad$ indices for general $\alpha$ orbitals \\ 
$\bar{p},\bar{q},\bar{r},\bar{s}$ &  $\qquad$ indices for general $\beta$ orbitals \\
$i,j,k,l$ &  $\qquad$ indices for $\alpha$ orbitals occupied in $|\bm{\mathrm{n}}\rangle$ \\ 
$\bar{i},\bar{j},\bar{k},\bar{l}$ &  $\qquad$ indices for $\beta$ orbitals occupied in $|\bm{\mathrm{n}}\rangle$ \\ 
$a,b,c,d$ &  $\qquad$ indices for $\alpha$ orbitals unoccupied in $|\bm{\mathrm{n}}\rangle$ \\ 
$\bar{a},\bar{b},\bar{c},\bar{d}$ &  $\qquad$ indices for $\beta$ orbitals unoccupied in $|\bm{\mathrm{n}}\rangle$ \\ 
%$J^{{}^{\alpha\alpha}}_{pq}$, $J^{{}^{\beta\beta}}_{\bar{p}\bar{q}}$, $J^{{}^{\alpha\beta}}_{p\bar{q}}$ &  $\qquad$ matrices of Jastrow variables \\
$J^{{}^{\alpha\alpha}}_{pq}$ &  $\qquad$ matrix of $\alpha$-$\alpha$ Jastrow variables \\
$J^{{}^{\beta\beta}}_{\bar{p}\bar{q}}$ &  $\qquad$ matrix of $\beta$-$\beta$ Jastrow variables \\
$J^{{}^{\alpha\beta}}_{p\bar{q}}$ &  $\qquad$ matrix of $\alpha$-$\beta$ Jastrow variables \\
$F_{p \bar{q}}$ &  $\qquad$ the full pairing matrix \\ 
$\Phi_{i \bar{j}}$ &  $\qquad$ matrix after deleting unoccupied rows \\
                   &  $\qquad$ and unoccupied columns from $\bm{F}$ \\ 
$R_{a \bar{i}}$    &  $\qquad$ matrix after deleting occupied rows \\
                   &  $\qquad$ and unoccupied columns from $\bm{F}$ \\
$C_{i \bar{a}}$    &  $\qquad$ matrix after deleting unoccupied rows \\
                   &  $\qquad$ and occupied columns from $\bm{F}$ \\
$B_{a \bar{b}}$    &  $\qquad$ matrix after deleting occupied rows \\
                   &  $\qquad$ and occupied columns from $\bm{F}$ \\
$\Theta_{\hspace{1pt}\bar{i} j}$ &  $\qquad$ inverse of the occupied pairing matrix $\bm{\Phi}$ \\ 
%$V_{i a j b}$ &  $\qquad$ $\alpha\alpha\alpha\alpha$ two-electron integrals in 1122 format \\ 
%$V_{i a \bar{j} \bar{b}}$ &  $\qquad$ $\alpha\alpha\beta\beta$ two-electron integrals in 1122 format \\ 
\hline\hline
\end{tabular}
\end{table}

\section{Theory}
\label{sec:theory}

\subsection{Basics}
\label{sec:basics}

In this paper we work with the JAGP ansatz in Hilbert space, given as
\begin{align}
&|\Psi\rangle = \hspace{1mm}\exp({\hat{J}}) |\Psi_{\mathrm{AGP}}\rangle
\label{eqn:jagp} \\
&|\Psi_{\mathrm{AGP}}\rangle = \hspace{1mm} \left( \sum_{r\bar{s}} F_{r\bar{s}} a^\dag_r a^\dag_{\bar{s}} \right)^{N/2} |0\rangle
\label{eqn:agp} \\
&\hat{J} = \sum_{p \le q}   J^{{}^{\alpha\alpha}}_{pq} \hat{n}_p\hat{n}_q
%\notag \\
  + \sum_{\bar{p} \le \bar{q}} J^{{}^{\beta\beta}}_{\bar{p}\bar{q}} \hat{n}_{\bar{p}}\hat{n}_{\bar{q}}
%\notag \\
  + \sum_{p\bar{q}} J^{{}^{\alpha\beta}}_{p\bar{q}} \hat{n}_p\hat{n}_{\bar{q}}
\label{eqn:jf} 
\end{align}
where $N/2$ is the number of $\alpha$ (and $\beta$) electrons, $\hat{n}_p$ and $a^\dag_p$ are the number and creation operators for orbital $p$,
$|0\rangle$ is the vacuum, and other symbols are as defined in Table \ref{tab:defs}.
The ``molecular'' geminal in Eq.\ (\ref{eqn:agp}) may contain a linear combination of any number of the non-orthogonal local geminals that
will be involved in different resonance structures, but rather than choosing them by hand we will allow the variational optimization of
the pairing matrix $\bm{F}$ to encode whatever geminal structure is most appropriate.
Also, note that the concise parameterization of the Jastrow factor in Eq.\ (\ref{eqn:jf}) is entirely equivalent to that used previously
\cite{Neuscamman:2012:sc_jagp} and maintains the ability to encode local particle number projection operators of the form
\begin{align}
\label{eqn:local_num_proj}
\exp\left( -\xi \left( M - \sum_{p \in W} \hat{n}_p \right)^2 \right)
\end{align}
where $\xi$ is the projection strength, $M$ is the desired local particle number, and $W$ is the set of orbitals defining the local region.
This equivalence ensures that the ansatz will remain size consistent.

To evaluate the energy of the JAGP, we use the variational energy expression
\begin{align}
E = \frac{\langle\Psi|H|\Psi\rangle}{\langle\Psi|\Psi\rangle} = \sum_{\bm{\mathrm{n}}} \frac{ |\langle\bm{\mathrm{n}}|\Psi\rangle|^2 }{\langle\Psi|\Psi\rangle }
\frac{\langle\bm{\mathrm{n}}|H|\Psi\rangle}{\langle\bm{\mathrm{n}}|\Psi\rangle},
\label{eqn:var_energy}
\end{align}
which may be evaluated by variational Monte Carlo (VMC) \cite{NigUmr-BOOK-99}
by sampling a set of occupations $\Omega$ from the wave function's probability 
distribution $|\langle\bm{\mathrm{n}}|\Psi\rangle|^2/\langle\Psi|\Psi\rangle$.
The energy is then estimated as an average of local energies,
\begin{align}
E = \frac{1}{n_s} \sum_{\bm{\mathrm{n}} \in \Omega}
\frac{\langle\bm{\mathrm{n}}|H|\Psi\rangle}{\langle\bm{\mathrm{n}}|\Psi\rangle} 
  \equiv \frac{1}{n_s} \sum_{\bm{\mathrm{n}} \in \Omega} E_L(\bm{\mathrm{n}}),
\label{eqn:vmc_energy}
\end{align}
with $n_s$ being the number of Monte Carlo samples in $\Omega$.
These samples are generated by a Metropolis-Hastings walk through Hilbert space, which requires us to evaluate wave function probability ratios between
different occupation number vectors.  If we use one-electron moves, then in order to correctly
accept or reject a proposed move that would transfer an electron from orbital $i$ to orbital $a$, we must evaluate ratios of the type
$\langle\bm{\mathrm{n}}|a^\dag_i a_a|\Psi\rangle/\langle\bm{\mathrm{n}}|\Psi\rangle$.
These ratios are also involved in computing local energies, and so we will assume for now that they can be evaluated efficiently,
a claim that will be verified over the course of the next few sections.

\subsection{Hamiltonian}
\label{sec:hamiltonian}

For our purposes, it is convenient to write the ab initio Hamiltonian as
\begin{align}
\label{eqn:ham}
&H = \hspace{2mm} \sum_{pq} t_{pq} \left( a^\dag_p a_q + a^\dag_{\bar{p}} a_{\bar{q}} \right) \\
&    \hspace{2mm} + \sum_{pqrs} (pq|rs) \left(  \frac{1}{2} a^\dag_p a_q a^\dag_r a_s
                                              + \frac{1}{2} a^\dag_{\bar{p}} a_{\bar{q}} a^\dag_{\bar{r}} a_{\bar{s}}
                                              +             a^\dag_p a_q a^\dag_{\bar{r}} a_{\bar{s}} \right),
\notag
\end{align}
where $(pq|rs)$ are the usual two-electron coulomb integrals \cite{Helgaker_book} in $(11|22)$ order
and $t_{pq}$ are modified one-electron integrals.
These modified integrals,
\begin{align}
\label{eqn:mod_oei}
t_{pq} = h_{pq} - \frac{1}{2} \sum_{r} (pr|rq),
\end{align}
where $h_{pq}$ are the standard one-electron integrals, are necessary to accommodate the ordering we have chosen for
the creation and destruction operators in Eq.\ (\ref{eqn:ham}).

Before considering our particular wave function, notice that the subset of Hamiltonian terms that contain only number
and hole operators (i.e.\ $\hat{n}_p=a^\dag_pa_p$ and $\hat{h}_p=a_pa^\dag_p$) will add a wave-function-independent contribution
to the local energy,
\begin{align}
E_0(\bm{\mathrm{n}}) =
\hphantom{+} & \sum_{i} t_{ii}
+ \sum_{\bar{i}} t_{\bar{i}\bar{i}}
+ \sum_{i \bar{j}} (i i|\bar{j} \bar{j})
\notag \\
+ & \frac{1}{2} \Bigg[ \hphantom{+}
    \sum_{i j} (i i|j j)
+   \sum_{i a} (i a|a i) 
\notag \\
\hphantom{+} & \hphantom{\frac{1}{2} \Bigg[}
+   \sum_{\bar{i} \bar{j}} (\bar{i} \bar{i}|\bar{j} \bar{j})
+   \sum_{\bar{i} \bar{a}} (\bar{i} \bar{a}|\bar{a} \bar{i}) 
\Bigg].
\label{eqn:ham_0}
\end{align}
Similarly, the two-electron components of $H$ that contain one but not two number or hole operators may be
converted in to one-electron operators by immediately acting the number or hole operator on $\langle\bm{\mathrm{n}}|$
in the numerator of Eq.\ (\ref{eqn:vmc_energy}).
The energy contribution of terms of this type can be accounted for through the one-electron expressions we
will derive below by using an $\bm{\mathrm{n}}$-specific modification of the one-electron integrals,
\begin{align}
t_{ia}^{\bm{\mathrm{n}}} =
t_{ia} & + \sum_j (ia|jj) + \sum_{\bar{j}} (ia|\bar{j}\bar{j}) \notag \\
       & + \frac{1}{2} \bigg[ \sum_b (ib|ba) - \sum_j (ja|ij) \bigg],
\label{eqn:tei_to_oei_aa} \\
t_{\bar{i} \bar{a}}^{\bm{\mathrm{n}}} =
t_{\bar{i} \bar{a}} & + \sum_{\bar{j}} (\bar{i} \bar{a}|\bar{j} \bar{j}) + \sum_j (\bar{i} \bar{a}|jj) \notag \\
        & + \frac{1}{2} \bigg[ \sum_{\bar{b}} (\bar{i} \bar{b}|\bar{b} \bar{a}) - \sum_{\bar{j}} (\bar{j} \bar{a}|\bar{i} \bar{j}) \bigg].
\label{eqn:tei_to_oei_bb}
\end{align}
With these modifications, the only Hamiltonian operators that remain to be dealt with are one-electron terms
of the type $a^\dag_i a_a$ and two-electron terms of the type $a^\dag_ia_aa^\dag_ja_b$, where $i\ne j$ and $a\ne b$,
which will be the focus of the next two sections.

\subsection{AGP energy}
\label{sec:agp_energy}

We will first show how the local AGP energy $\langle\bm{\mathrm{n}}|H|\Psi_{\mathrm{AGP}}\rangle/\langle\bm{\mathrm{n}}|\Psi_{\mathrm{AGP}}\rangle$
can be calculated efficiently before showing in the next section how the effects of Jastrow factors may be incorporated by an efficient
modification of the one- and two-electron integrals.
To begin our treatment of the AGP local energy, we recall that the AGP's coefficient for a given occupation number vector is equal
to a determinant of the occupied portion of the pairing matrix \cite{Bouchaud:1988:agp_det},
\begin{align}
\label{eqn:agp_coeff}
\langle\bm{\mathrm{n}}|\Psi_{\mathrm{AGP}}\rangle = \mathrm{det}(\bm{\Phi}).
\end{align}
Thus ratios of the type $\langle\bm{\mathrm{n}}|a^\dag_i a_a|\Psi_{\mathrm{AGP}}\rangle/\langle\bm{\mathrm{n}}|\Psi_{\mathrm{AGP}}\rangle$ that
arise in computing the one-electron component of the local energy are simply ratios of determinants that differ by one row (i.e.\ row $i$ has
been replaced by row $a$ in the numerator's determinant).
Such ratios can be efficiently computed without evaluating the determinants themselves by using the ``rank $m$'' matrix determinant
lemma \cite{MatrixRefManual}, which states that for any invertible $n\times n$ matrix $\bm{A}$
\begin{align}
\label{eqn:mat_det_lemma}
\mathrm{det}\left(\bm{A}+\bm{U}\bm{V}^\intercal\right) = \mathrm{det}\left(\bm{I} + \bm{V}^\intercal\bm{A}^{-1}\bm{U}\right)
                                                         \mathrm{det}\left(\bm{A}\right)
\end{align}
for any $n\times m$ matrices $\bm{U}$ and $\bm{V}$, where $\bm{I}$ is the $m\times m$ identity matrix.
To evaluate the ratios we need, we construct $\bm{U}$ and $\bm{V}$ such that they replace row $i$ in the determinant
$\langle\bm{\mathrm{n}}|\Psi_{\mathrm{AGP}}\rangle$
with its counterpart in the determinant
$\langle\bm{\mathrm{n}}|a^\dag_i a_a|\Psi_{\mathrm{AGP}}\rangle$.
This replacement is executed via the rank 1 update defined by
\begin{align}
\label{eqn:one_e_uv}
U_{k1} = \delta_{ki} \qquad \qquad V_{\bar{k}1} = R_{a\bar{k}} - \Phi_{i\bar{k}}
\end{align}
where $\bm{U}$ picks out the row to replace, $\bm{V}$ holds the difference between the new and old rows, and
$\delta_{ki}$ is the Kronecker delta.
Using these definitions, Eq.\ (\ref{eqn:mat_det_lemma}) allows us to write the ratio we desire as
\begin{align}
\frac{\langle\bm{\mathrm{n}}|a^\dag_i a_a|\Psi_{\mathrm{AGP}}\rangle}{\langle\bm{\mathrm{n}}|\Psi_{\mathrm{AGP}}\rangle}
&= 1 + \sum_{\bar{k} l} \left( R_{a\bar{k}} - \Phi_{i\bar{k}} \right) \Theta_{\bar{k}l} \delta_{li} \notag \\
&= ( \bm{R} \bm{\Theta} )_{ai}
\label{eqn:singly_excited_ratio}
\end{align}
where the second equality follows because $\bm{\Theta}$ is the inverse of $\bm{\Phi}$.
Recognizing that the $\beta\beta$ part of the one-electron energy will lead to an analogous formula involving
single column replacements, we see that the one-electron part of the local energy may be evaluated as
\begin{align}
\label{eqn:1e_energy}
E^{\mathrm{AGP}}_1 =   \sum_{ia} t_{ia}^{\bm{\mathrm{n}}} ( \bm{R} \bm{\Theta} )_{ai}
                     + \sum_{\bar{i}\bar{a}} t_{\bar{i}\bar{a}}^{\bm{\mathrm{n}}} ( \bm{\Theta} \bm{C} )_{\bar{i}\bar{a}}
\end{align}
where $t_{ia}^{\bm{\mathrm{n}}}$ and  $t_{\bar{i}\bar{a}}^{\bm{\mathrm{n}}}$ are as defined in
Eqs.\ (\ref{eqn:tei_to_oei_aa}-\ref{eqn:tei_to_oei_bb}).
%with a cost that scales as $n_s n_o^2 n_u$ due to the matrix multiplications $\bm{R} \bm{\Theta}$ and $\bm{\Theta} \bm{C}$.

Note that we need not perform direct matrix inversion to compute $\bm{\Theta}$ at each step in our random walk.
Rather, we first compute $\bm{\Theta}$ directly for the occupation number vector that initiates the walk, after which
we update it using the Sherman-Morrison formula each time we make a move.
The cost of this updating procedure for $\bm{\Theta}$ scales as $n_s n_o^2$, which is less than direct inversion's
$n_s n_o^3$ and much less than the $n_s n_o^2 n_u^2$ that will be required for the overall energy.

For the two-electron component of the Hamiltonian, we will encounter ratios between determinants whose matrices
differ by two rows, two columns, or one row and one column, and so we will employ the rank 2 matrix determinant lemma.
For the all-$\alpha$-spin two-electron part, the energy contribution is
\begin{align}
E^{\mathrm{AGP}}_{2\alpha\alpha} &=
\frac{1}{2} \sum_{i a j b} (i a | j b) \frac{\langle\bm{\mathrm{n}}|a^\dag_i a_a a^\dag_j a_b|\Psi_{\mathrm{AGP}}\rangle}
                                            {\langle\bm{\mathrm{n}}|\Psi_{\mathrm{AGP}}\rangle} \notag \\
&\equiv \frac{1}{2} \sum_{i a j b} (i a | j b) \mathcal{R}^{ab}_{ij}
\label{eqn:aaaa_tei_energy}
\end{align}
where $\mathcal{R}^{ab}_{ij}$ is a ratio in which the numerator determinant $\langle\bm{\mathrm{n}}|a^\dag_i a_a a^\dag_j a_b|\Psi_{\mathrm{AGP}}\rangle$
has substituted rows $a$ and $b$ for rows $i$ and $j$ relative to $\langle\bm{\mathrm{n}}|\Psi_{\mathrm{AGP}}\rangle$.
%(Note that here we ignore terms from the two-electron portion of Eq.\ (\ref{eqn:ham}) in which $p=q$, $p=s$, $r=q$, or $r=s$ as such terms can
%be trivially included in the one-electron energy of Eq.\ (\ref{eqn:1e_energy}) by a further occupation-number-vector-specific modification of the
%one-electron integrals).
This double row substitution can be written using the rank 2 update
\begin{alignat}{3}
U_{k1} &= \delta_{ki} && \qquad \qquad & V_{\bar{k}1} &= R_{a\bar{k}} - \Phi_{i\bar{k}} \notag \\
U_{k2} &= \delta_{kj} && \qquad \qquad & V_{\bar{k}2} &= R_{b\bar{k}} - \Phi_{j\bar{k}}
\label{eqn:aaaa_uv}
\end{alignat}
which when used with Eq.\ (\ref{eqn:mat_det_lemma}) allows us to write the ratios from
Eq.\ (\ref{eqn:aaaa_tei_energy}) as 2$\times$2 determinants,
%\begin{align}
%\mathcal{R}^{ab}_{ij}
%& = \begin{vmatrix}
%\hspace{2mm}  1 + \sum_{\bar{k}} \left(R_{a \bar{k}} - \Phi_{i \bar{k}}\right) \Theta_{\bar{k} i} \hspace{1mm} &
%\hspace{8mm}      \sum_{\bar{k}} \left(R_{a \bar{k}} - \Phi_{i \bar{k}}\right) \Theta_{\bar{k} j} \hspace{2mm} \\
%\hspace{8mm}      \sum_{\bar{k}} \left(R_{b \bar{k}} - \Phi_{j \bar{k}}\right) \Theta_{\bar{k} i} \hspace{1mm} &
%\hspace{2mm}  1 + \sum_{\bar{k}} \left(R_{b \bar{k}} - \Phi_{j \bar{k}}\right) \Theta_{\bar{k} j} \hspace{2mm} \end{vmatrix}.
%\label{eqn:aaaa_det_ratio}
%\end{align}
%By again making use of the definition $\bm{\Theta}=\bm{\Phi}^{-1}$ we may simplify Eq.\ (\ref{eqn:aaaa_det_ratio}) to 
\begin{align}
\mathcal{R}^{ab}_{ij}
& = \begin{vmatrix}
\left(\bm{R} \bm{\Theta} \right)_{ai} &
\left(\bm{R} \bm{\Theta} \right)_{aj} \\
\left(\bm{R} \bm{\Theta} \right)_{bi} &
\left(\bm{R} \bm{\Theta} \right)_{bj} \end{vmatrix}.
\label{eqn:aaaa_det_ratio_simp}
\end{align}
We expand this determinant to form the final expression for the AGP's all-$\alpha$-spin contribution to the two-electron energy,
\begin{align}
\label{eqn:aaaa_tei_expanded}
& E^{\mathrm{AGP}}_{2\alpha\alpha} = \\
& \hspace{3mm} \frac{1}{2} \sum_{i a j b} (i a | j b)
\bigg [ ~
  (\bm{R}\bm{\Theta})_{ai} (\bm{R}\bm{\Theta})_{bj}
- (\bm{R}\bm{\Theta})_{aj} (\bm{R}\bm{\Theta})_{bi}
~ \bigg ].
\notag
\end{align}
The expression for the all-$\beta$-spin component is derived analogously via column replacements, resulting in
\begin{align}
\label{eqn:bbbb_tei_expanded}
& E^{\mathrm{AGP}}_{2\beta\beta} = \\
& \hspace{3mm} \frac{1}{2} \sum_{\bar{i} \bar{a} \bar{j} \bar{b}} (\bar{i} \bar{a} | \bar{j} \bar{b})
\bigg [ ~
  (\bm{\Theta}\bm{C})_{\bar{i} \bar{a}} (\bm{\Theta}\bm{C})_{\bar{j} \bar{b}}
- (\bm{\Theta}\bm{C})_{\bar{j} \bar{a}} (\bm{\Theta}\bm{C})_{\bar{i} \bar{b}}
~ \bigg ].
\notag
\end{align}
Finally, the $\alpha\beta$ two-electron contribution,
\begin{align}
\label{eqn:aabb_tei_energy}
E^{\mathrm{AGP}}_{2\alpha\beta} =
\sum_{i a \bar{j} \bar{b}} (i a | \bar{j} \bar{b}) \frac{\langle\bm{\mathrm{n}}|a^\dag_i a_a a^\dag_{\bar{j}} a_{\bar{b}}|\Psi_{\mathrm{AGP}}\rangle}
                                            {\langle\bm{\mathrm{n}}|\Psi_{\mathrm{AGP}}\rangle}
\end{align}
involves ratios in which row $i$ is replaced with row $a$ and column $\bar{j}$ is replaced with column $\bar{b}$.
This slightly more complicated case can be dealt with via the rank 2 update
\begin{alignat}{3}
U_{k1} &= \delta_{ki}                   && \qquad \qquad & V_{\bar{k}1} &= (1-\delta_{\bar{k}\bar{j}})R_{a\bar{k}} + \delta_{\bar{k}\bar{j}}B_{a\bar{b}} - \Phi_{i\bar{k}} \notag \\
V_{\bar{k}2} &= \delta_{\bar{k}\bar{j}} && \qquad \qquad & U_{k2} &= (C_{k\bar{b}} - \Phi_{k\bar{j}})(1-\delta_{ki})
\label{eqn:aabb_uv}
\end{alignat}
where we have taken care that the element where the row and column cross is handled correctly.
This rank 2 update allows us to use the matrix determinant lemma to write
(after some algebra similar to but messier than the development in Eqs.\ (\ref{eqn:aaaa_uv}-\ref{eqn:aaaa_tei_expanded}))
the $\alpha\beta$ two-electron contribution as
\begin{align}
\label{eqn:aabb_tei_expanded}
& E^{\mathrm{AGP}}_{2\alpha\beta} = \\
& \hspace{3mm} \sum_{i a \bar{j} \bar{b}} (i a | \bar{j} \bar{b})
\bigg[ (\bm{R}\bm{\Theta})_{ai} (\bm{\Theta} \bm{C})_{\bar{j}\bar{b}}
+ \Theta_{\bar{j}i} \big( \bm{B} - \bm{R} \bm{\Theta} \bm{C} \big)_{a\bar{b}} \bigg].
\notag
\end{align}

Thus, by repeated use of the matrix determinant lemma, we are able to write the bare AGP's local energy as a sum of the
contributions in Eqs.\ (\ref{eqn:ham_0}), (\ref{eqn:1e_energy}), (\ref{eqn:aaaa_tei_expanded}), (\ref{eqn:bbbb_tei_expanded}), and
(\ref{eqn:aabb_tei_expanded}), each of which consists of efficient contractions of the one- and two-electron integrals
with different subsections of the pairing matrix $\bm{F}$ and the inverse $\bm{\Theta}$ of its occupied part.
An inspection of these contributions reveals that the overall cost to compute the AGP's energy in this way scales 
polynomially as $n_s n_o^2 n_u^2$.
In the next section, we will show that this scaling remains unchanged upon adding the Jastrow factor.

\subsection{Jastrow AGP energy}
\label{sec:jagp_energy}

Having constructed an efficient energy evaluation for the AGP wave function, it remains to add the effects
of the Jastrow factor, with which the local energy expression becomes
\begin{align}
E_L(\bm{\mathrm{n}}) = \frac{\langle\bm{\mathrm{n}}|H \mathrm{exp}(\hat{J})|\Psi_{\mathrm{AGP}}\rangle}
                            {\langle\bm{\mathrm{n}}|  \mathrm{exp}(\hat{J})|\Psi_{\mathrm{AGP}}\rangle}.
\label{eqn:jf_local_energy}
\end{align}
As before, we will begin by handling the one-electron components of $H$, after which an analogous but
slightly more intricate approach will be developed for the two-electron components.

In incorporating effects of the Jastrow factor on the $\alpha$-spin one-electron terms of the local energy,
\begin{align}
\label{eqn:jagp_1e_le}
E^{\mathrm{JAGP}}_{1\alpha} = \sum_{ia} t_{ia}^{\bm{\mathrm{n}}} 
\frac{\langle\bm{\mathrm{n}}|a^\dag_i a_a \mathrm{exp}(\hat{J})|\Psi_{\mathrm{AGP}}\rangle}
     {\langle\bm{\mathrm{n}}|\mathrm{exp}(\hat{J})|\Psi_{\mathrm{AGP}}\rangle},
\end{align}
we will again need to evaluate ratios between wave function coefficients for occupation number vectors
differing by one orbital.
These ratios are no longer simply the ratios of determinants, but we may still take advantage of the
determinantal properties used in the previous section by first converting the Jastrow factors into scalars.
To see how this is done, consider the denominator.
By acting the Jastrow factor $\mathrm{exp}(\hat{J})$ to the left, we may convert the number operators
$\hat{n}_p$ and $\hat{n}_{\bar{p}}$ inside it into the occupation numbers $n_p$ and $n_{\bar{p}}$ of
$\langle\bm{\mathrm{n}}|$.
At this point, the whole Jastrow factor becomes a scalar, which we will write as
$\mathrm{exp}(J(\bm{\mathrm{n}}))$, and can be factored out of the expectation value.
Doing the same for the numerator and defining $\bm{\mathrm{n}}_i^a$ as the result of replacing orbital
$i$ with orbital $a$ in $\bm{\mathrm{n}}$, the ratio in Eq.\ (\ref{eqn:jagp_1e_le}) becomes
the product of a ratio of Jastrow factors and the determinantal ratio we encountered in the previous section,
\begin{align}
\label{eqn:jagp_1e_le_factored}
E^{\mathrm{JAGP}}_{1\alpha} = \sum_{ia} t_{ia}^{\bm{\mathrm{n}}} 
\frac{\mathrm{exp}(J(\bm{\mathrm{n}}_i^a))}{\mathrm{exp}(J(\bm{\mathrm{n}}))}
\frac{\langle\bm{\mathrm{n}}|a^\dag_i a_a |\Psi_{\mathrm{AGP}}\rangle}
     {\langle\bm{\mathrm{n}}|\Psi_{\mathrm{AGP}}\rangle}.
\end{align}
We can now simplify the ratio of Jastrows by recognizing that all components of the numerator and
denominator Jastrows cancel out except those that touch orbitals $i$ or $a$.
Defining the ($\bm{\mathrm{n}}$-dependent) intermediates
\begin{align}
\label{eqn:jastrow_single_intermed}
K^\alpha_p =
J^{\alpha\alpha}_{pp} + \sum_{k\ne p} J^{\alpha\alpha}_{pk} + \sum_{\bar{k}} J^{\alpha\beta}_{p\bar{k}}
\end{align}
and the Jastrow-transformed one-electron integrals
\begin{align}
\tilde{t}_{ia}^{\bm{\mathrm{n}}} & = t_{ia}^{\bm{\mathrm{n}}} \hspace{1mm} \mathrm{exp}\big( J(\bm{\mathrm{n}}_i^a) - J(\bm{\mathrm{n}}) \big) \notag \\
                                 & = t_{ia}^{\bm{\mathrm{n}}} \hspace{1mm} \mathrm{exp}\big( K^\alpha_a - K^\alpha_i - J^{\alpha\alpha}_{ia} \big)
\label{eqn:jastrow_trans_oei}
\end{align}
we may rewrite our energy expression as
\begin{align}
E^{\mathrm{JAGP}}_{1\alpha} & = \sum_{ia} \tilde{t}_{ia}^{\bm{\mathrm{n}}} 
\frac{\langle\bm{\mathrm{n}}|a^\dag_i a_a |\Psi_{\mathrm{AGP}}\rangle}
     {\langle\bm{\mathrm{n}}|\Psi_{\mathrm{AGP}}\rangle}
 = \sum_{ia} \tilde{t}_{ia}^{\bm{\mathrm{n}}} ( \bm{R} \bm{\Theta} )_{ai}
\label{eqn:jagp_1e_le_final}
\end{align}
where the second equality follows from Eq.\ (\ref{eqn:singly_excited_ratio}).

Using the same strategy, we can modify the two electron integrals to account for the Jastrow factor's effect
on the two-electron component of the local energy.
The all-$\alpha$ two-electron contribution can be computed by replacing $(ia|jb)$ in Eq.\ (\ref{eqn:aaaa_tei_expanded})
with the transformed integrals
\begin{align}
[ia|jb] & \equiv (ia|jb) \hspace{1mm} \mathrm{exp}\big( J(\bm{\mathrm{n}}_{ij}^{ab}) - J(\bm{\mathrm{n}}) \big) \notag \\
        & = (ia|jb) \hspace{1mm} \mathrm{exp}\bigg[ \hspace{2mm} K^\alpha_a + K^\alpha_b - K^\alpha_i - K^\alpha_j \notag \\
        & \hphantom{= (ia|jb) \hspace{1mm} \mathrm{exp}\bigg[} \hspace{1mm} + J^{\alpha\alpha}_{ab} - J^{\alpha\alpha}_{ia} - J^{\alpha\alpha}_{jb} \notag \\
        & \hphantom{= (ia|jb) \hspace{1mm} \mathrm{exp}\bigg[} \hspace{1mm} + J^{\alpha\alpha}_{ij} - J^{\alpha\alpha}_{ib} - J^{\alpha\alpha}_{ja} \hspace{2mm} \bigg],
%        & = (ia|jb) \hspace{1mm} \mathrm{exp}\bigg[ \hspace{1mm} \hphantom{+} K^\alpha_a - J^{\alpha\alpha}_{ia} - J^{\alpha\alpha}_{ja} \notag \\
%        & \hphantom{= (ia|jb) \hspace{1mm} \mathrm{exp}\bigg[} \hspace{1mm} + K^\alpha_b - J^{\alpha\alpha}_{ib} - J^{\alpha\alpha}_{jb} \notag \\
%        & \hphantom{= (ia|jb) \hspace{1mm} \mathrm{exp}\bigg[} \hspace{1mm} + J^{\alpha\alpha}_{ab} + J^{\alpha\alpha}_{ij} - K^\alpha_i - K^\alpha_j \bigg],
\label{eqn:jastrow_trans_tei_aaaa}
\end{align}
while the all-$\beta$ contribution is computed using the analogous replacement in Eq.\ (\ref{eqn:bbbb_tei_expanded}).
Similarly, the $\alpha\beta$ component may be evaluated by replacing $(ia|\bar{j}\bar{b})$ in Eq.\ (\ref{eqn:aabb_tei_expanded})
with the transformed integrals
\begin{align}
[ia|\bar{j}\bar{b}] & \equiv (ia|\bar{j}\bar{b}) \hspace{1mm} \mathrm{exp}\big( J(\bm{\mathrm{n}}_{i\bar{j}}^{a\bar{b}}) - J(\bm{\mathrm{n}}) \big) \notag \\
        & = (ia|\bar{j}\bar{b}) \hspace{1mm} \mathrm{exp}\bigg[ \hspace{2mm} K^\alpha_a + K^\beta_{\bar{b}} - K^\alpha_i - K^\beta_{\bar{j}} \notag \\
        & \hphantom{= (ia|\bar{j}\bar{b}) \hspace{1mm} \mathrm{exp}\bigg[} \hspace{1mm} + J^{\alpha\beta}_{a\bar{b}} - J^{\alpha\alpha}_{ia} - J^{\beta\beta}_{\bar{j}\bar{b}} \notag \\
        & \hphantom{= (ia|\bar{j}\bar{b}) \hspace{1mm} \mathrm{exp}\bigg[} \hspace{1mm} + J^{\alpha\beta}_{i\bar{j}} - J^{\alpha\beta}_{i\bar{b}} - J^{\alpha\beta}_{a\bar{j}} \hspace{2mm} \bigg].
\label{eqn:jastrow_trans_tei_aabb}
\end{align}
Putting all these developments together, we see that the JAGP local energy is
\begin{align}
& E_L(\bm{\mathrm{n}}) = \hspace{2mm}
   E_0(\bm{\mathrm{n}})
 + \sum_{ia} \tilde{t}_{ia}^{\bm{\mathrm{n}}} ( \bm{R} \bm{\Theta} )_{ai}
 + \sum_{\bar{i} \bar{a}} \tilde{t}_{\bar{i} \bar{a}}^{\bm{\mathrm{n}}} ( \bm{\Theta} \bm{C} )_{\bar{i} \bar{a}}
 \notag \\
& \hspace{3mm} + \frac{1}{2} \sum_{i a j b} [i a | j b]
\bigg [ ~
  (\bm{R}\bm{\Theta})_{ai} (\bm{R}\bm{\Theta})_{bj}
- (\bm{R}\bm{\Theta})_{aj} (\bm{R}\bm{\Theta})_{bi}
~ \bigg ]
\notag \\
& \hspace{3mm} + \frac{1}{2} \sum_{\bar{i} \bar{a} \bar{j} \bar{b}} [\bar{i} \bar{a} | \bar{j} \bar{b}]
\bigg [ ~
  (\bm{\Theta}\bm{C})_{\bar{i} \bar{a}} (\bm{\Theta}\bm{C})_{\bar{j} \bar{b}}
- (\bm{\Theta}\bm{C})_{\bar{j} \bar{a}} (\bm{\Theta}\bm{C})_{\bar{i} \bar{b}}
~ \bigg ] \notag \\
& \hspace{3mm} + \sum_{i a \bar{j} \bar{b}} [i a | \bar{j} \bar{b}]
\bigg[ (\bm{R}\bm{\Theta})_{ai} (\bm{\Theta} \bm{C})_{\bar{j}\bar{b}}
+ \Theta_{\bar{j}i} \big( \bm{B} - \bm{R} \bm{\Theta} \bm{C} \big)_{a\bar{b}} \bigg],
\label{eqn:final_jagp_le}
\end{align}
and that the total energy of Eq.\ (\ref{eqn:vmc_energy}) can be evaluated for a cost that scales as $n_s n_o^2 n_u^2$.
Thus with $n_s$ growing linearly with system size to control statistical uncertainties, the cost to
evaluate the JAGP energy grows polynomially as the fifth power of system size.

\subsection{Derivative ratios}
\label{sec:derivatives}

Before discussing the variational minimization of the JAGP energy, we first lay the groundwork by
developing efficient evaluations of terms that we call derivative ratios, which will make the optimization
relatively simple to describe.
Using the compact derivative notation $|\Psi^x\rangle\equiv\partial|\Psi\rangle/\partial \mu_x$, where $\mu_x$ is the
$x$th wave function parameter, we first consider the ``bare'' derivative ratio
\begin{align}
\mathcal{D}_{\bm{\mathrm{n}}}(\mu_x) \equiv \frac{\langle\bm{\mathrm{n}}|\Psi^x\rangle}{\langle\bm{\mathrm{n}}|\Psi\rangle}.
\label{eqn:bare_dr_definition}
\end{align}
%which is the ratio of $\langle\bm{\mathrm{n}}|\Psi^x\rangle$, the coefficient of the wave function's derivative with respect to the parameter $\mu_x$ to
%the wave function coefficient $\langle\bm{\mathrm{n}}|\Psi\rangle$ on the configuration $\bm{\mathrm{n}}$.
For Jastrow variables, these ratios are quite trivial,
\begin{align}
\mathcal{D}_{\bm{\mathrm{n}}}(J^{\alpha\alpha}_{pq}) &= n_p n_q \label{eqn:jf_aa_bare_dr} \\
\mathcal{D}_{\bm{\mathrm{n}}}(J^{\beta\beta}_{\bar{p}\bar{q}}) &= n_{\bar{p}} n_{\bar{q}} \label{eqn:jf_bb_bare_dr} \\
\mathcal{D}_{\bm{\mathrm{n}}}(J^{\alpha\beta}_{p\bar{q}}) &= n_p n_{\bar{q}} \label{eqn:jf_ab_bare_dr}
\end{align}
as a derivative of the exponential in Eq.\ (\ref{eqn:jagp}) simply brings down the pair of number operators associated
with the Jastrow variable in question.
For pairing matrix variables, we see from Eq.\ (\ref{eqn:agp_coeff}) that only derivatives with respect to the occupied portion $\bm{\Phi}$ will be
non-zero and that these may be evaluated via the properties of determinant derivatives,
\begin{align}
\mathcal{D}_{\bm{\mathrm{n}}}(\Phi_{i\bar{j}}) &= \frac{1}{\mathrm{det}(\bm{\Phi})} \frac{\partial \mathrm{det}(\bm{\Phi})}{\partial \Phi_{i\bar{j}}}
= \Theta_{\bar{j}i}.
\label{eqn:pm_bare_dr}
\end{align}

In addition to the bare derivative ratios, we also define the energy derivative ratios,
\begin{align}
\mathcal{G}_{\bm{\mathrm{n}}}(\mu_x) \equiv \frac{\langle\bm{\mathrm{n}}|H|\Psi^x\rangle}{\langle\bm{\mathrm{n}}|\Psi\rangle},
\label{eqn:energy_dr_definition}
\end{align}
which are related to derivatives of the local energy by
\begin{align}
\mathcal{G}_{\bm{\mathrm{n}}}(\mu_x) = \frac{\partial E_L(\bm{\mathrm{n}})}{\partial \mu_x} + \mathcal{D}_{\bm{\mathrm{n}}}(\mu_x) E_L(\bm{\mathrm{n}}).
\label{eqn:loc_en_der}
\end{align}
We now assert that for a given $\bm{\mathrm{n}}$, the full set of energy derivative ratios can be
evaluated analytically at a cost that scales as $n_o^2 n_u^2$.
To see how, consider that the local energy given in Eq.\ (\ref{eqn:final_jagp_le}) can be constructed from $\bm{F}$,
$\bm{J}^{\alpha\alpha}$, $\bm{J}^{\beta\beta}$, $\bm{J}^{\alpha\beta}$, and $\bm{\Theta}$ using only addition,
multiplication, and exponentiation.
Furthermore, derivatives of $\bm{\Theta}=\bm{\Phi}^{-1}$ with respect to the elements of $\bm{\Phi}$ may be
evaluated using the properties of the matrix inverse,
\begin{align}
\frac{\partial \Theta_{\bar{j}i}}{\partial \Phi_{k\bar{l}}} = - \Theta_{\bar{j}k} \Theta_{\bar{l}i}.
\label{eqn:mat_inv_der}
\end{align}
Thus the local energy $E_L(\bm{\mathrm{n}})$ is a single-output, continuously differentiable function of the
inputs $\bm{F}$, $\bm{J}^{\alpha\alpha}$, $\bm{J}^{\beta\beta}$, and $\bm{J}^{\alpha\beta}$, and so by the
theory of algorithmic differentiation \cite{Griewank-Walther-book}, all of its analytic derivatives with respect to these inputs
can be computed for an overall cost that is a small constant multiple of the cost of evaluating
$E_L(\bm{\mathrm{n}})$ itself.
We therefore see that the full set of local energy derivatives, and, by
Eq.\ (\ref{eqn:loc_en_der}), energy derivative ratios, can be evaluated at a cost that scales as $n_s n_o^2 n_u^2$.
Although we will omit the detailed equations for $\mathcal{G}_{\bm{\mathrm{n}}}(\mu_x)$ here, they can be
derived by a straightforward application of adjoint algorithmic differentiation \cite{Griewank-Walther-book} to
Eq.\ (\ref{eqn:final_jagp_le}).

\subsection{Optimization}
\label{sec:optimization}

We variationally optimize the energy of the JAGP wave function using the linear method (LM)
\cite{UmrTouFilSorHen-PRL-07,TouUmr-JCP-08},
which we will briefly review here.
The LM works by repeatedly solving the Schr\"{o}dinger equation in a special subspace of the full Hilbert
space defined by the span of the wave function and its first derivatives.
More precisely, we construct the generalized eigenvalue problem
\begin{align}
\sum_{y\in\{0,1,\ldots\}} \langle\Psi^x| H |\Psi^y \rangle c_y = \lambda \sum_{y\in\{0,1,\ldots\}} \langle\Psi^x|\Psi^y \rangle c_y
\label{eqn:gen_eig_eqn}
\end{align}
where $|\Psi^x\rangle$ and $|\Psi^y\rangle$ are shorthand for the derivatives of $|\Psi\rangle$
with respect to the $x$th and $y$th wave function parameters $\mu_x$ and $\mu_y$, respectively,
and $|\Psi^0\rangle\equiv|\Psi\rangle$.
After solving this eigenvalue problem for $\bm{c}$, one updates the parameters by
\begin{align}
\mu_x \leftarrow \mu_x + c_x / c_0 \quad\qquad x\in\{1,2,\ldots\},
\label{eqn:parameter_update}
\end{align}
after which the updated $|\Psi\rangle$ will be a good approximation for the subspace eigenfunction
defined by $\bm{c}$ so long as the updates $c_x/c_0$ are not too large.
%This optimization scheme has proved effective at optimizing large sets of nonlinear wave function
%parameters in the past, and in the results below we will see that it is significantly more
%effective than the quasi-Newton BFGS (cite) optimization.

In order to perform the LM optimization, we must solve Eq.\ (\ref{eqn:gen_eig_eqn}), which requires
building or at least multiplying by the Hamiltonian and overlap matrices.
To see how either of these operations may be accomplished using the derivative ratios presented in
the previous section, we insert resolutions of the identity
$\sum_{\bm{\mathrm{n}}}|\bm{\mathrm{n}}\rangle\langle\bm{\mathrm{n}}|$
on both sides, divide by $\langle\Psi|\Psi\rangle$, and multiply and divide by the square of the local
wave function coefficient $|\langle\bm{\mathrm{n}}|\Psi\rangle|^2$ to obtain
\begin{align}
& \sum_{\bm{\mathrm{n}}} \sum_{y\in\{0,1,\ldots\}}
\frac{|\langle\bm{\mathrm{n}}|\Psi\rangle|^2}
     {\langle\Psi|\Psi\rangle}
\frac{\langle\Psi^x|\bm{\mathrm{n}}\rangle}
     {\langle\Psi  |\bm{\mathrm{n}}\rangle}
\frac{\langle\bm{\mathrm{n}}| H |\Psi^y \rangle}
     {\langle\bm{\mathrm{n}}|    \Psi   \rangle}
c_y \notag \\
& \hspace{6mm} = \lambda
\sum_{\bm{\mathrm{n}}} \sum_{y\in\{0,1,\ldots\}}
\frac{|\langle\bm{\mathrm{n}}|\Psi\rangle|^2}
     {\langle\Psi|\Psi\rangle}
\frac{\langle\Psi^x|\bm{\mathrm{n}}\rangle}
     {\langle\Psi  |\bm{\mathrm{n}}\rangle}
\frac{\langle\bm{\mathrm{n}}|\Psi^y \rangle}
     {\langle\bm{\mathrm{n}}|\Psi   \rangle}
c_y
\label{eqn:gen_eig_res_identity}
\end{align}
which can be approximated on the Monte Carlo sample set $\Omega$ as
\begin{align}
& \sum_{\bm{\mathrm{n}}\in\Omega} \sum_{y\in\{0,1,\ldots\}}
\frac{\langle\Psi^x|\bm{\mathrm{n}}\rangle}
     {\langle\Psi  |\bm{\mathrm{n}}\rangle}
\frac{\langle\bm{\mathrm{n}}| H |\Psi^y \rangle}
     {\langle\bm{\mathrm{n}}|    \Psi   \rangle}
c_y \notag \\
& \hspace{8mm} = \lambda
\sum_{\bm{\mathrm{n}}\in\Omega} \sum_{y\in\{0,1,\ldots\}}
\frac{\langle\Psi^x|\bm{\mathrm{n}}\rangle}
     {\langle\Psi  |\bm{\mathrm{n}}\rangle}
\frac{\langle\bm{\mathrm{n}}|\Psi^y \rangle}
     {\langle\bm{\mathrm{n}}|\Psi   \rangle}
c_y.
\label{eqn:gen_eig_qmc}
\end{align}
Inspecting Eq.\ (\ref{eqn:gen_eig_qmc}) makes clear that the Monte Carlo approximations to 
the Hamiltonian and overlap matrices in the first derivative subspace can either be
multiplied by or built explicitly using only the derivative ratios
$\mathcal{D}_{\bm{\mathrm{n}}}(\mu_x)$ and $\mathcal{G}_{\bm{\mathrm{n}}}(\mu_x)$ from
Eqs.\ (\ref{eqn:bare_dr_definition}) and (\ref{eqn:energy_dr_definition}), and so
we see that the LM can therefore be applied efficiently by evaluating these ratios
as outlined in the previous section.

\section{Results}
\label{sec:results}

Before we present our results, let us briefly overview the computational details.
JAGP calculations used one- and two-electron integrals generated by Psi3 \cite{Psi3},
with the atomic orbital basis orthonormalized via the $S^{-1/2}$ Lowdin procedure.
For all methods, we have frozen the 1s core orbital in non-hydrogen elements.
Data for full configuration interaction (FCI) \cite{Handy:1984:fci,Handy:1989:fci},
complete active space perturbation theory (CASPT2) \cite{Werner:1996:caspt2},
Davidson-corrected multi-reference configuration interaction (MRCI+Q)
\cite{Knowles:1988:mrci,Werner:1988:mrci},
and restricted open shell (ROHF) methods
were computed using \uppercase{MOLPRO} \cite{MOLPRO_brief}.
Data for second order M{\o}ller-Plesset perturbation theory (MP2 and UMP2)
\cite{Szabo-Ostland}, unrestricted perfect pairing (UPP) \cite{Beran:2005:upp},
B3LYP \cite{BECKE:1993:b3lyp},
and coupled cluster with singles, doubles, and perturbative triples (CCSD(T) and UCCSD(T))
\cite{BARTLETT:2007:cc_review} were computed using QChem \cite{QChem:2013}.

\subsection{Convergence of linear method}
\label{sec:convergence}

\begin{figure}[t]
\centering
\includegraphics[width=7.5cm,angle=270]{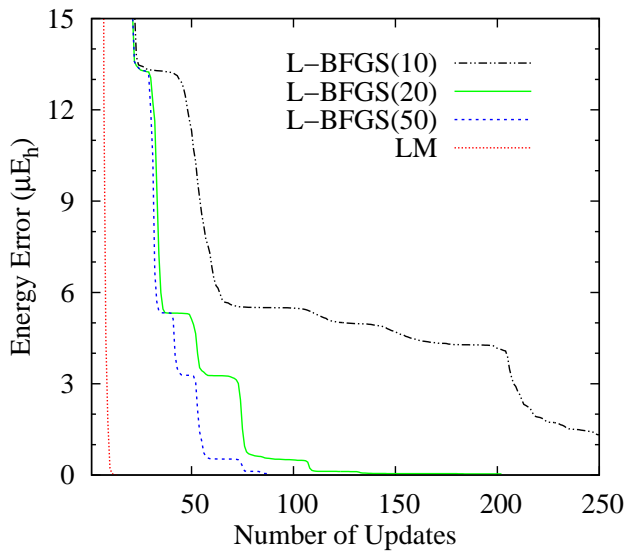}
\caption{Convergence of the JAGP energy error vs FCI for two nearby H$_2$ molecules in the STO-3G
         basis for different optimization methods, with the numbers in parentheses being the
         L-BFGS history lengths.
         %See Section \ref{sec:convergence}.
        }
\label{fig:h4_sto3g_conv_energy}
\end{figure}

We begin our results by looking at two examples illustrating the convergence properties
of the linear method optimization.
In the first example, we perform an energy optimization on two neighboring H$_2$ molecules
(with atom xyz positions in Angstroms
[0.0,  0.0,  0.0],
[1.0,  0.0,  0.0],
[0.2,  3.0, -0.1], and
[0.2,  3.0,  0.9])
in a minimal STO-3G basis \cite{Pople:1969:sto-3g}.
To remove any confusion caused by statistical uncertainty in Monte Carlo sampling, we have
executed the optimizations for this example on an ``exact'' sample, in which every
occupation number vector is visited once and given a weight of
$|\langle\bm{\mathrm{n}}|\Psi\rangle|^2$.
Thus here we are concerned only with the quality of the LM optimization in an ideal case.
The next example will show concerns that arise during normal Monte Carlo sampling.

\begin{figure}[t]
\centering
\includegraphics[width=7.5cm,angle=270]{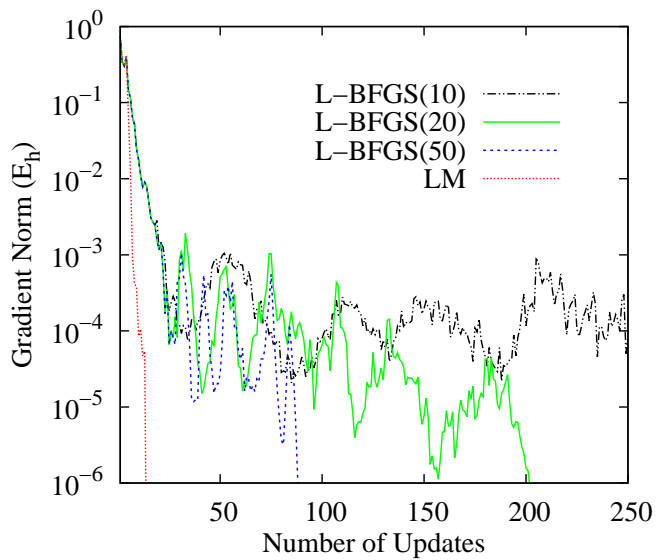}
\caption{Convergence of the JAGP energy gradient for two nearby H$_2$ molecules in the STO-3G
         basis for different optimization methods, with the numbers in parentheses being the
         L-BFGS history lengths.
         %See Section \ref{sec:convergence}.
        }
\label{fig:h4_sto3g_conv_gradient}
\end{figure}

\begin{figure}[b]
\centering
\includegraphics[width=7.5cm,angle=270]{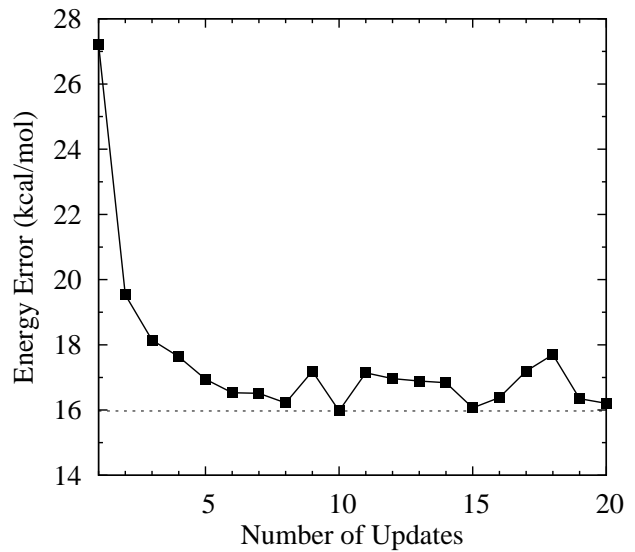}
\caption{Convergence of the JAGP energy error vs FCI for N$_2$ at $r_{{}_{\mathrm{NN}}}=1.4$\ \AA\ 
         in the 6-31G basis using the linear method, with the initial guess being the
         optimized wave function for $r_{{}_{\mathrm{NN}}}=1.5$\ \AA.
         Statistical uncertainties for the individual points are smaller than the symbol size,
         and the dashed line shows our best JAGP energy for this geometry.
         %See Section \ref{sec:convergence}.
        }
\label{fig:n2_631g_conv_energy}
\end{figure}

In this first example we compare the effectiveness of the LM with that of the low-memory BFGS
algorithm (L-BFGS) \cite{Nocedal:1980:lbfgs}, which performs a quasi-Newton minimization with
a Hessian approximated by finite differences among a stored history of energy gradients.
This optimization is simple to execute for our wave function, as it requires only the energy
and its gradients with respect to the wave function parameters, which may be easily constructed
using the derivative ratios of Sec.\ \ref{sec:derivatives}.
In Figures \ref{fig:h4_sto3g_conv_energy} and \ref{fig:h4_sto3g_conv_gradient},
we see that the LM greatly outperforms the L-BFGS algorithm in both lowering the energy
and reducing the energy gradient, converging in a small fraction of the number of iterations
necessary for L-BFGS.
Note that even in the first 50 iterations, during which the length-50 history L-BFGS is
equivalent to full BFGS, the LM is still far superior.
We also note that, thanks to the size-consistency recovered by the Jastrow factor,
the JAGP is an essentially exact description of this system.
This accuracy should not be surprising, as the system is constructed of weakly interacting
two-electron subsystems, an ideal case for JAGP.

\begin{figure}[t]
\centering
\includegraphics[width=6.0cm,angle=0]{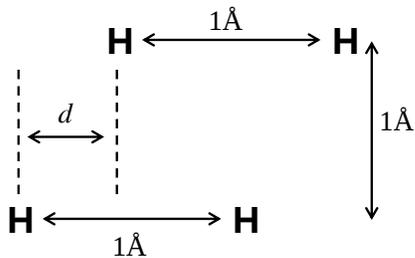}
\caption{Schematic of the H$_4$ distortion, where the geometry is a square
         when the offset $d=0$.
         %See Section \ref{sec:hydrogen}.
        }
\label{fig:hydrogen_diagram}
\end{figure}

\begin{figure}[b]
\centering
\includegraphics[width=7.5cm,angle=270]{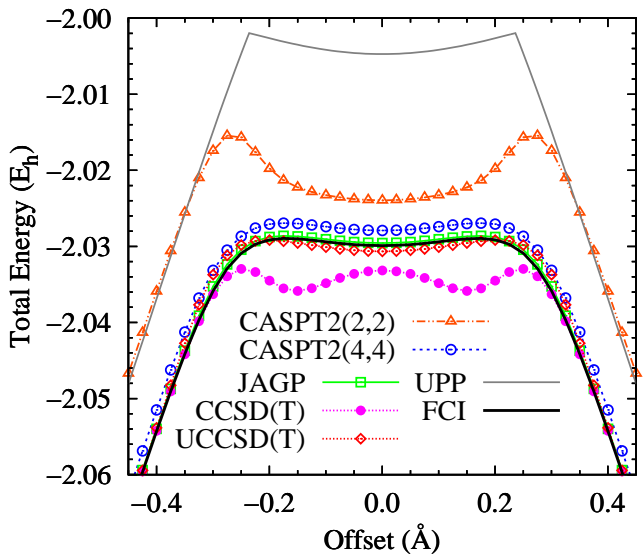}
\caption{Total energies of the H$_4$ distortion of Figure \ref{fig:hydrogen_diagram} for different offsets $d$.
         The 6-31G basis set was used, and statistical uncertainties for JAGP are smaller than the symbol size.
         %See Section \ref{sec:hydrogen}.
        }
\label{fig:h4_631g_energy}
\end{figure}

For our second example, we inspect a typical Monte Carlo optimization for the
nitrogen molecule in the 6-31G basis set \cite{POPLE:1972:6-31g_basis}.
Two features of the optimization shown in Figure \ref{fig:n2_631g_conv_energy} stand
out.
First, we see a rapid convergence to the energy minimum, with fewer than ten iterations
necessary to get within 1 kcal/mol of the best JAGP energy.
Second, we see that as the minimum is approached, the energies of successive
updates stop decreasing monotonically, with oscillations observed that are outside the
statistical uncertainties of the individual energy points.
%These oscillations indicate that while a LM update of course lowers the energy on
%the current random sample of occupation number vectors, 
These oscillations are caused by updates in which the parameters are easily
``over-optimized'' for a particular random sample in a way that
lowers the energy for that sample while raising it for other samples.
While we can prevent this over-optimization by taking larger sample sets,
such effort is not strictly necessary.
%These oscillations appear to be due to over-optimization on a given Monte Carlo sample,
%as the estimates for the updated energies given by the LM eigenvalues in
%Eq.\ (\ref{eqn:gen_eig_eqn}) are in this region invariably lower than the energies computed
%on a fresh Monte Carlo sample using the updated wave function (the latter being what
%is plotted).
%As one would expect from this hypothesis, these oscillations are reduced when larger samples
%are taken.
%Such extra effort is not strictly necessary, however, as we will show
Indeed, we will see in the remainder of the results that the minimum energies produced by
these somewhat noisy optimizations are already quite accurate, and due to the variational principle
we do not have to worry about these oscillations causing us to overshoot the exact energy.

\begin{figure}[b]
\centering
\includegraphics[width=7.5cm,angle=270]{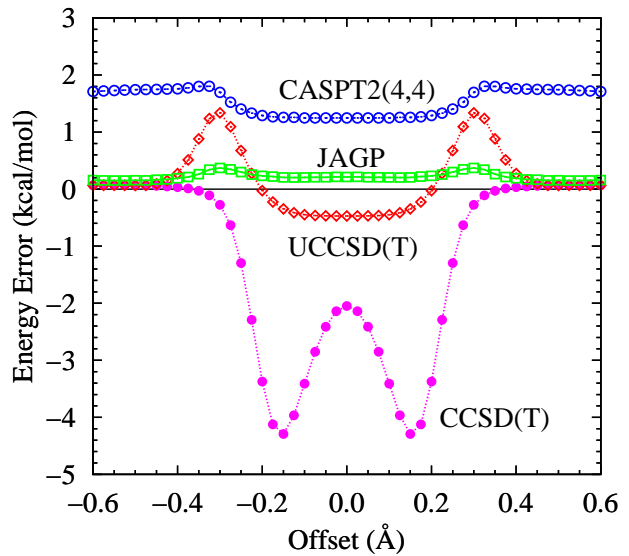}
\caption{Energy errors vs FCI for the H$_4$ distortion of Figure \ref{fig:hydrogen_diagram} for different offsets $d$.
         The 6-31G basis set was used, and statistical uncertainties for JAGP are smaller than the symbol size.
         %See Section \ref{sec:hydrogen}.
        }
\label{fig:h4_631g_error}
\end{figure}

\subsection{H$_4$ distortion}
\label{sec:hydrogen}

To demonstrate the JAGP's ability to handle strong correlations between more than
two electrons in a simple and clear example, we have applied it to the four-hydrogen
distortion coordinate shown in Figure \ref{fig:hydrogen_diagram}, which deforms a square of
hydrogen atoms into a parallelogram.
While the strongest correlations in this system arise from the two higher-energy electrons
that must navigate the near-degeneracy of the second and third molecular orbitals, significant strong
correlations are present for all four electrons, as evidenced by the fact that a (2,2) active space
is an insufficient starting point for CASPT2.
Indeed, we see in Figure \ref{fig:h4_631g_energy} that CASPT2 is only correct when starting with a
(4,4) ``full valence'' active space, and so by the definition that correlation is strong when it
cannot be treated perturbatively, we see that all four electrons are strongly correlated.

As one might expect from this active space analysis, the high-level single-reference method CCSD(T)
fails qualitatively in this system, predicting unphysical local minima at offsets of $d=\pm 0.15$\ \AA.
These unphysical minima can be removed by breaking spin symmetry in the Hartree-Fock reference
and generalizing to UCCSD(T).
However, this approach simply trades one qualitative failure for another, as the UHF determinant is
strongly spin contaminated with an unphysical $\langle S^2\rangle\approx 1$ for small offsets.
Even with this sacrifice, we see in Figure \ref{fig:h4_631g_error} that UCCSD(T) is still not
energetically competitive with true multi-reference methods.

\begin{figure}[t]
\centering
\includegraphics[width=7.5cm,angle=270]{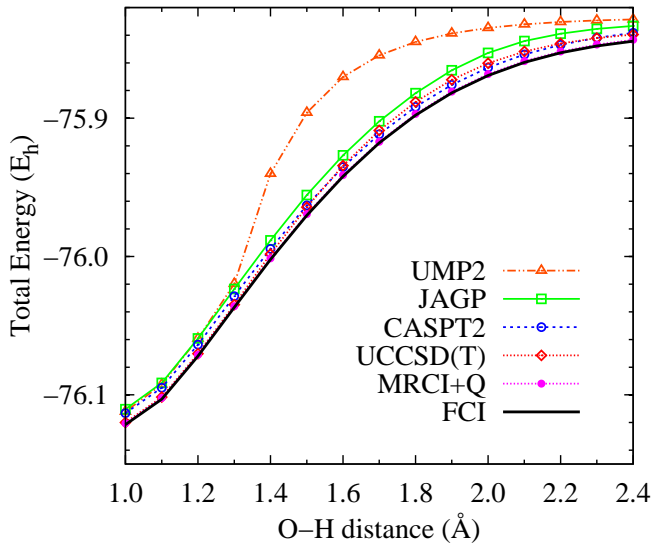}
\caption{Total energies for the symmetric stretch of H$_2$O in the 6-31G basis with
         a bond angle of 109.57$^\circ$.
         Statistical uncertainties for JAGP are smaller than the symbol size.
         %See Section \ref{sec:h2o}.
        }
\label{fig:h2o_631g_energy}
\end{figure}

In this example we see clearly that the JAGP is capable of treating strong correlation even when more
than two electrons are involved, a fundamental step beyond the capabilities of PP.
Restricted PP (not shown) fails even to predict an energy plateau, while unrestricted
PP (UPP) displays sharp cusps in its potential energy surface.
These cusps are caused by the crossing of UPP solutions \cite{Parkhill:2009:perfect_quads} that
represent different valence bond structures.
JAGP, on the other hand, is able to create a superposition of these competing valence bond structures, removing
the unphysical cusps and producing a good spin eigenstate with $\langle S^2\rangle < 0.01$ at all offsets.
This feature of JAGP appears to be relatively robust:
JAGP predicts pure spin eigenstates in all systems tested, even when spin unrestriction is necessary to
produce a qualitatively correct coupled cluster result.
In terms of energy error, JAGP is the most accurate method tested in this system in both an absolute sense
and in terms of non-parallelity error (NPE), defined as the difference between a method's largest and smallest
deviations from FCI.
%with accuracies (as measured by
%non-parallelity error) better than coupled cluster and approaching CASPT2.

%Notably, the JAGP ansatz achieves its high accuracy in H$_4$ without breaking spin symmetry or relying on a
%``full valence'' (and hence exponentially costly) active space reference.
%As we will see in the following sections, this robustness appears to be a relatively general feature
%of JAGP:  in all systems tested in which spin unrestriction is necessary to create a qualitatively
%correct coupled cluster result, JAGP predicts pure spin eigenstates with accuracies (as measured by
%non-parallelity error) better than coupled cluster and approaching CASPT2.
%Furthermore, it does this using QMC algorithms that do not suffer from the exponential scaling
%problem of active space methods.

\subsection{H$_2$O stretch}
\label{sec:h2o}

\begin{figure}[t]
\centering
\includegraphics[width=7.5cm,angle=270]{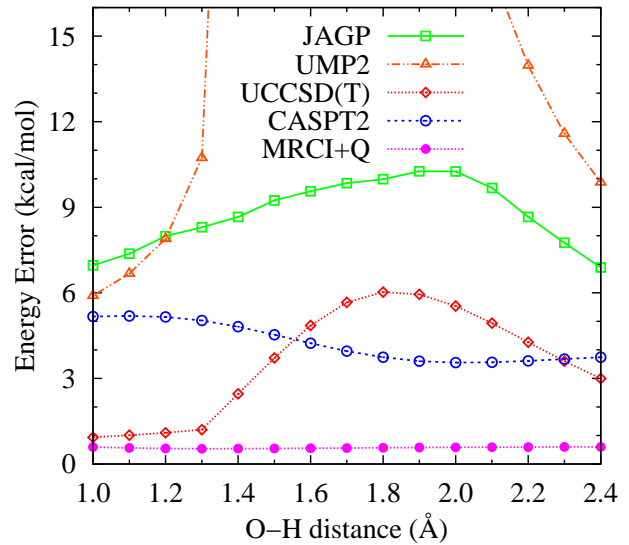}
\caption{Energy errors vs FCI for the symmetric stretch of H$_2$O in the 6-31G basis.
         Statistical uncertainties for JAGP are smaller than the symbol size.
         %See Section \ref{sec:h2o}.
        }
\label{fig:h2o_631g_error}
\end{figure}

Previously \cite{Neuscamman:2012:sc_jagp} we reported results on the symmetric H$_2$O stretch in a minimal
STO-3G basis, showing that JAGP's worst error relative to FCI was less than 2 kcal/mol.
Here we extend this demonstration to include dynamic correlation, showing results for the symmetric stretch
in a 6-31G basis in Figures \ref{fig:h2o_631g_energy} and \ref{fig:h2o_631g_error}.
Two features are immediately clear.
First, JAGP still delivers a qualitatively correct potential energy surface with a 3.4 kcal/mol NPE
that is smaller than the 5.1 kcal/mol NPE of UCCSD(T) (restricted CCSD(T) performs extremely
poorly and is not shown).
Second, absolute errors are significantly larger in 6-31G as compared to STO-3G, topping out around
10 kcal/mol at an OH distance of 2 \AA.
A third important feature not shown in the figures is that the optimized JAGP wave function is a good
spin eigenstate with $\langle S^2\rangle$ calculated to be less than 0.01 across the entire stretching
coordinate.

Given JAGP's high accuracy in capturing the strong correlations present in STO-3G, these findings imply
that it lacks a full accounting of the details of dynamic correlation.
This shortcoming is most obvious near equilibrium, where JAGP is missing 7 kcal/mol of correlation energy
even though strong correlations are not present.
Clearly, future development of this theory should focus on capturing the remaining dynamic correlation,
a point to which we will return in the conclusion.
Nonetheless, for the H$_2$O stretch, the JAGP provides a more accurate potential energy surface than
unrestricted single reference methods while avoiding the exponential cost of active space methods.

\begin{figure}[t]
\centering
\includegraphics[width=7.5cm,angle=270]{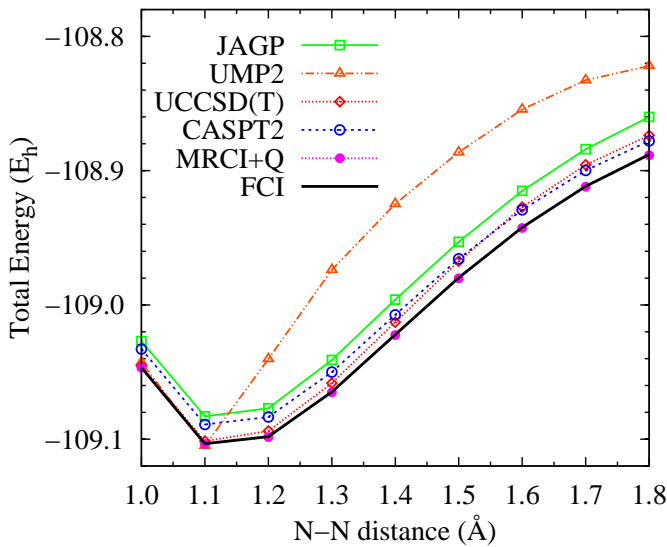}
\caption{Total energies for stretched N$_2$ in the 6-31G basis.
         Statistical uncertainties for JAGP are smaller than the symbol size.
         %See Section \ref{sec:n2}.
        }
\label{fig:n2_631g_tot_energy}
\end{figure}

\subsection{N$_2$ stretch}
\label{sec:n2}

\begin{figure}[t]
\centering
\includegraphics[width=7.5cm,angle=270]{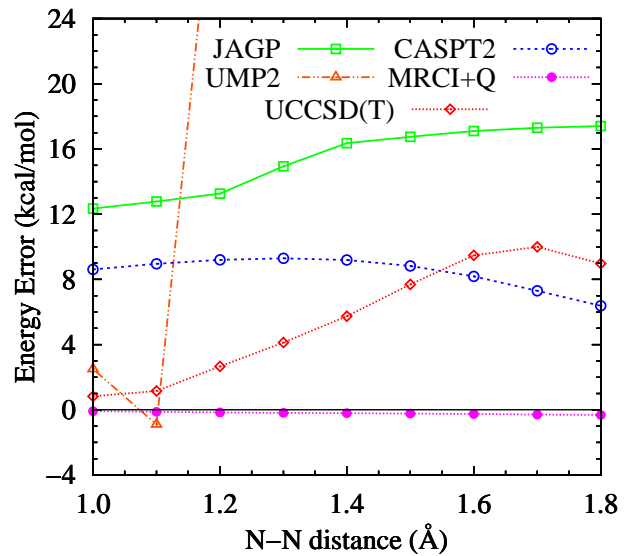}
\caption{Energy errors vs FCI for stretched N$_2$ in the 6-31G basis.
         Statistical uncertainties for JAGP are smaller than the symbol size.
         %See Section \ref{sec:n2}.
        }
\label{fig:n2_631g_error}
\end{figure}

JAGP results for stretching the N$_2$ triple bond are shown in Figures \ref{fig:n2_631g_tot_energy}
and \ref{fig:n2_631g_error}.
As in H$_2$O, the NPE is better than that of single reference methods but not as good as
that of CASPT2, while absolute errors are relatively large (12 to 17 kcal/mol).
Again, the state is a good spin singlet across the entire stretching coordinate.

Although at first glance these results do not seem to provide much insight beyond what was gleaned
from H$_2$O, it is important to consider that in N$_2$ the six strongly correlated electrons are
all in close proximity to each other, while in H$_2$O there is at least some spatial separation
into two distinct bonds.
We therefore conclude that while the JAGP is ideally suited for describing spatially separated pairs,
%of strongly correlating electrons,
it is not wholly reliant on such structure and can produce faithful descriptions of the strong
correlations of even a close-proximity six-electron problem like N$_2$.
%This insight again highlights JAGP's advantage over PP theories, whose strongly orthogonal
%geminals cannot account for the inter-pair correlations that are present in a close-proximity
%six-electron problem like N$_2$.

\subsection{C$_2$ triplet}
\label{sec:c2}

\begin{figure}[t]
\centering
\includegraphics[width=7.5cm,angle=270]{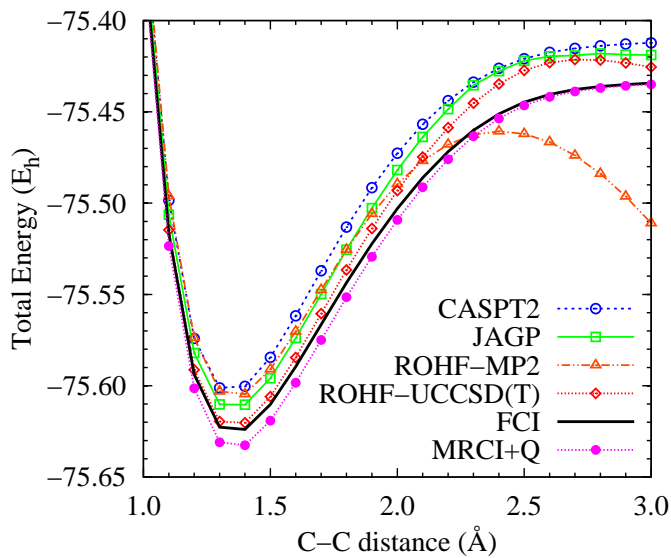}
\caption{Total energies for the triplet B${}_{\mathrm{2u}}$ state of C$_2$
         in the 6-31G basis.
         Statistical uncertainties for JAGP are smaller than the symbol size.
         %See Section \ref{sec:c2}.
        }
\label{fig:c2_631g_tot_energy}
\end{figure}

\begin{figure}[t]
\centering
\includegraphics[width=7.5cm,angle=270]{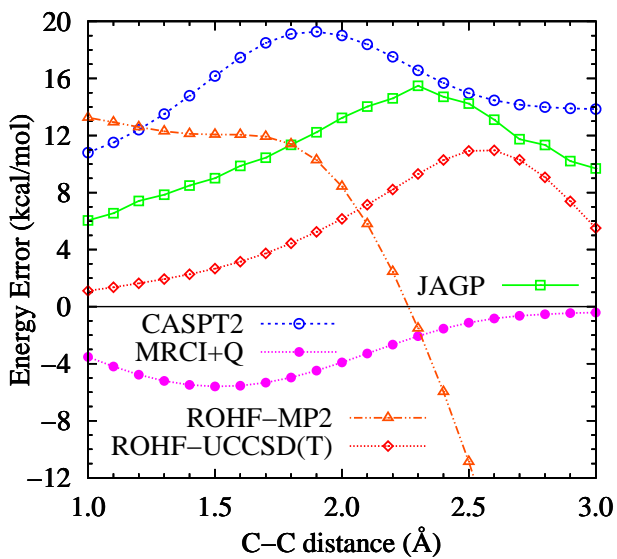}
\caption{Energy errors vs FCI for the triplet B${}_{\mathrm{2u}}$ state of C$_2$
         in the 6-31G basis.
         Statistical uncertainties for JAGP are smaller than the symbol size.
         %See Section \ref{sec:c2}.
        }
\label{fig:c2_631g_error}
\end{figure}

To test the JAGP's performance in a non-singlet state, we have applied it to the triplet
B${}_{\mathrm{2u}}$ state of C$_2$, for which stretching coordinate data are shown in
Figures \ref{fig:c2_631g_tot_energy} and \ref{fig:c2_631g_error}.
We accessed the triplet state by adding a distant H$_2$ molecule to the system and applying
number projecting Jastrow factors to ensure the net $S_z$ spin of C$_2$ was +1 while that
of H$_2$ was -1 (thus circumventing the restriction that an AGP wave function must have equal
numbers of $\alpha$ and $\beta$ electrons).
In our results we have subtracted away the energy of the H$_2$ triplet.
The B${}_{\mathrm{2u}}$ representation was enforced by using symmetrized atomic orbitals and
allowing the random walk to visit only those configurations with B${}_{\mathrm{2u}}$
symmetry.
This scheme is essentially another way to perform variation after symmetry and number projection
\cite{Scuseria:2012:projected_hf} with the added benefit of including Jastrow factors.

We see in the results that all methods have more trouble in C$_2$ than for
H$_2$O or N$_2$, with JAGP giving a non-parallelity error slightly better than that of
ROHF-UCCSD(T) and slightly worse than that of CASPT2.
Notably, even MRCI+Q has difficulty in this example, showing a variational violation and
an uncharacteristically large non-parallelity error.
Overall, we see that JAGP's performance for a triplet state is similar to that for singlets,
delivering a good spin eigenstate ($\langle S^2\rangle = 2$) and an accuracy approaching
that of CASPT2.
%This is encouraging as it indicates the method should be more widely applicable than it's
%singlet-based history suggests.

\subsection{Ethene}
\label{sec:ethene}

\begin{figure}[t]
\centering
\includegraphics[width=6.0cm,angle=0]{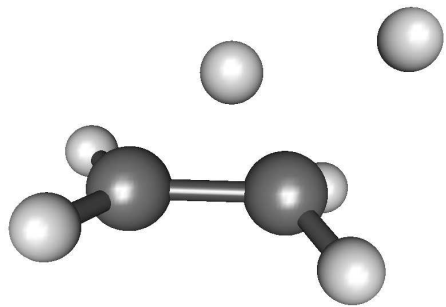}
\caption{Transition state for the C$_s$ mechanism.
         %See Section \ref{sec:ethene}.
        }
\label{fig:cs_ts}
\end{figure}

\begin{figure}[t]
\centering
\includegraphics[width=6.0cm,angle=0]{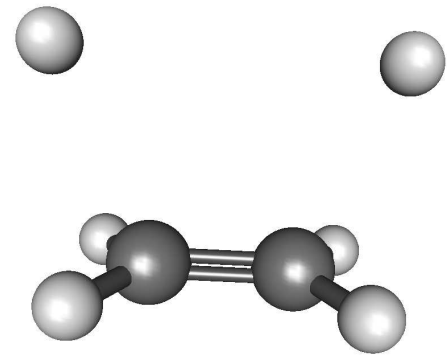}
\caption{Transition state for the C$_{2v}$ mechanism.
         %See Section \ref{sec:ethene}.
        }
\label{fig:c2v_ts}
\end{figure}

As a final example, we have used the JAGP wave function to predict barrier heights
for the unassisted gas phase insertion of H$_2$ into ethene's double bond.
Using CASSCF with a (4,4) active space in the 6-31G basis, we have located the two
saddle point structures shown in Figures \ref{fig:cs_ts} and \ref{fig:c2v_ts} using the quadratic steepest
descent (QSD) method \cite{Ruedenberg:1993:qsd_ts}
as implemented in MOLPRO (coordinates can be found in the Appendix).
The first structure has $\mathrm{C}_s$ symmetry, is single-reference in nature, and has
been found previously using density functional theory \cite{Radom:2000:tm_free_hydrogenation}.
The second structure has $\mathrm{C}_{2v}$ symmetry, is multi-reference in nature, and
does not appear to have been reported previously.

These saddle points correspond to transition states for two different mechanisms
of the unassisted gas phase hydrogenation reaction.
We have verified that they are indeed transition states by computing intrinsic reaction
coordinates (IRCs) towards both products and reactants starting from these structures
(again via the QSD method).
In both cases, these structures are the maximum-energy points for their respective IRCs.
The mechanism for the $\mathrm{C}_s$ transition state involves ``sliding'' H$_2$ in
from one side, which has the advantage of not having to fully break the H$_2$ bond
until late in the reaction coordinate when significant C-H bonding has already begun.
In the $\mathrm{C}_{2v}$ mechanism, on the other hand, the H$_2$ bond is broken already
at the transition structure, which at first glance makes this mechanism appear uncompetitive.

What is especially interesting about the $\mathrm{C}_{2v}$ structure is that despite
its multi-reference nature (the CASSCF active natural orbital occupations are 1.9, 1.2, 0.8, and 0.1),
and the fact that the H$_2$ bond is already broken, its barrier height is essentially the
same as that of the single-reference $\mathrm{C}_{s}$ structure.
Indeed, JAGP predicts the $\mathrm{C}_{s}$ and $\mathrm{C}_{2v}$ barriers to be
136 and 135 kcal/mol, respectively, as shown in Table \ref{tab:ethene_barriers}.
This nearly equal-height prediction is corroborated by the benchmark MRCI+Q and by UCCSD(T),
but \textit{not} by CCSD(T), MP2, UMP2, B3LYP, or UB3LYP, which predict to different degrees a significant
difference between the two barriers.
%Note that spin-unrestricted Hartree-Fock and B3LYP do not lower the restricted determinant's energy
%at either transition state.
Thus at temperatures high enough for this unassisted reaction to proceed,
high-level methods predict that two mechanistic channels will be in essentially equal competition,
while other methods (especially MP2, UB3LYP, and B3LYP) erroneously predict
that one pathway will be significantly favored.

\section{Conclusions}
\label{sec:conclusions}

This report has detailed efficient algorithms for evaluating and optimizing the energy of the
Jastrow antisymmetric geminal power (JAGP) wave function in Hilbert space.
This ansatz represents an alternative approach for approximating the electronic structure of
molecules and materials that relies on neither the independent particle approximation nor
the active space framework.
Instead, the basic assumption is that the strongest correlations between electrons can be
broken down in a pairwise fashion, with the overall wave function constructed as a superposition
of valence bonding structures built in turn from products of local two-electron functions.
The JAGP attempts to encode this structure by placing all electrons in a single ``molecular''
geminal that holds an arbitrary linear combination of the non-orthogonal local geminal
building blocks.
Raising this geminal to the appropriate power produces a superposition of geminal products,
from which the Jastrow factor selects the desired resonating valence bond structures
by projecting out terms with undesirable local particle number distributions.

%\begin{figure*}[h]
%\centering
%\begin{minipage}{.5\textwidth}
%  \centering
%  \includegraphics[width=6cm]{sr_ts}
%  %\captionof{figure}{A figure}
%  \label{fig:cs_ts}
%\end{minipage}%
%\begin{minipage}{.5\textwidth}
%  \centering
%  \includegraphics[width=6cm]{mr_ts}
%  %\captionof{figure}{Another figure}
%  \label{fig:c2v_ts}
%\end{minipage}
%\caption{C$_s$ (left) and C$_{2v}$ (right) transition states.  See Section \ref{sec:ethene}.}
%\label{fig:ethene_ts}
%\end{figure*}

Our results demonstrate that for a variety of strongly correlated examples, JAGP delivers good
spin eigenstates with potential energy surfaces superior in quality to those of the most
accurate single-reference methods and approaching the quality of CASPT2 results.
These examples include singlet and triplet multi-bond dissociations as well as a reaction
mechanism involving a novel multi-reference transition state.
Furthermore, this accuracy is coupled with good formal properties:  the JAGP is variational,
size consistent, and can be optimized
in polynomial time for a cost that scales as the fifth power of the system size.
%In contrast, every other method discussed in this paper either scales exponentially,
%lacks one or more of these
%properties or fails to produce even qualitatively correct results for strongly
%correlated systems.

\begin{table}[t]
\centering
\caption{Barrier heights (kcal/mol) for ethene hydrogenation mechanisms at the
         6-31G CASSCF(4,4) geometries.
         JAGP statistical uncertainties are less than 1 kcal/mol.
         %See Section \ref{sec:ethene}.
        }
\label{tab:ethene_barriers}
\begin{tabular}{  c   r@{.}l   r@{.}l  }
\hline\hline
\hspace{0mm} Method \hspace{0mm} &
\multicolumn{2}{ c }{ \hspace{0mm} $\mathrm{C}_{s}$   \hspace{0mm} } &
\multicolumn{2}{ c }{ \hspace{0mm} $\mathrm{C}_{2v}$  \hspace{0mm} } \\
\hline
%                                                      Cs                                 C2v
 \hspace{0mm} B3LYP     \hspace{0mm}  &  \hspace{0mm} 130&0  \hspace{0mm} & \hspace{0mm} 170&3 \hspace{0mm} \\
 \hspace{0mm} UB3LYP    \hspace{0mm}  &  \hspace{0mm} 130&0  \hspace{0mm} & \hspace{0mm} 146&4 \hspace{0mm} \\
 \hspace{0mm} MP2       \hspace{0mm}  &  \hspace{0mm} 139&8  \hspace{0mm} & \hspace{0mm} 177&5 \hspace{0mm} \\
 \hspace{0mm} UMP2      \hspace{0mm}  &  \hspace{0mm} 144&4  \hspace{0mm} & \hspace{0mm} 139&9 \hspace{0mm} \\
 \hspace{0mm} CCSD(T)   \hspace{0mm}  &  \hspace{0mm} 131&0  \hspace{0mm} & \hspace{0mm} 125&5 \hspace{0mm} \\
 \hspace{0mm} UCCSD(T)  \hspace{0mm}  &  \hspace{0mm} 132&8  \hspace{0mm} & \hspace{0mm} 134&6 \hspace{0mm} \\
 \hspace{0mm} MRCI+Q    \hspace{0mm}  &  \hspace{0mm} 130&8  \hspace{0mm} & \hspace{0mm} 132&5 \hspace{0mm} \\
 \hspace{0mm} JAGP      \hspace{0mm}  &  \hspace{0mm} 135&9  \hspace{0mm} & \hspace{0mm} 135&2 \hspace{0mm} \\
\hline\hline
\end{tabular}
\end{table}

The most pressing question that remains to be answered for JAGP is whether its ability to
describe dynamic correlation can be improved.
One strategy is to use JAGP as a guiding function in fixed-node projector Monte Carlo
\cite{Ceperley:1995:ub_fixed_node}, which is the Hilbert space analogue of fixed-node diffusion
Monte Carlo (DMC) \cite{FouMitNeeRaj-RMP-01} and should be capable of capturing the remaining
dynamic correlation.
Alternatively, if the local-particle-number-projecting qualities of the Hilbert space Jastrow
factor can be transported into real space, then a combination of traditional cusp-providing
Jastrow factors \cite{FouMitNeeRaj-RMP-01} and real space DMC would likely serve to provide
the missing dynamic correlation.
Finally, even within the VMC framework, there is still scope to improve the JAGP by
generalizing to a Pfaffian geminal \cite{Bajdich:2006:pfaffian,Bajdich:2008:pfaffian} and by
optimizing the local orbital basis in which the Jastrow factor is defined.
Research in these exciting directions is underway.

\section{Acknowledgments}
\label{sec:acknowledgments}

We thank Martin Head-Gordon for many insightful discussions and for computational resources.
We thank the Miller Institute for Basic Research in Science for funding.

\appendix
\section{Transition States}
\label{sec:appendix}

\begin{table}[h]
\centering
\caption{Geometry, in Angstroms, of the CASSCF(4,4) 6-31G C$_s$ transition state of the unassisted gas phase ethene hydrogenation reaction.
        }
\label{tab:cs_ts_struct}
\begin{tabular}{  c   r@{.}l   r@{.}l  r@{.}l  }
\hline\hline
 C    &  \hspace{2mm}   0&0000000000    & \hspace{2mm}  0&0087063743   & \hspace{2mm}  -0&7990056481 \\
 C    &  \hspace{2mm}   0&0000000000    & \hspace{2mm} -0&1355804403   & \hspace{2mm}   0&6628393017 \\
 H    &  \hspace{2mm}   0&0000000000    & \hspace{2mm}  1&0510492173   & \hspace{2mm}   0&2326305655 \\
 H    &  \hspace{2mm}   0&0000000000    & \hspace{2mm}  1&2421143685   & \hspace{2mm}   1&8005988762 \\
 H    &  \hspace{2mm}   0&9012117308    & \hspace{2mm} -0&4823664902   & \hspace{2mm}   1&1231614214 \\
 H    &  \hspace{2mm}  -0&9012117308    & \hspace{2mm} -0&4823664902   & \hspace{2mm}   1&1231614214 \\
 H    &  \hspace{2mm}  -0&9255826314    & \hspace{2mm}  0&0917247367   & \hspace{2mm}  -1&3284709121 \\
 H    &  \hspace{2mm}   0&9255826314    & \hspace{2mm}  0&0917247367   & \hspace{2mm}  -1&3284709121 \\
\hline\hline
\end{tabular}
\end{table}

\begin{table}[h]
\centering
\caption{Geometry, in Angstroms, of the CASSCF(4,4) 6-31G C$_{2v}$ transition state of the unassisted gas phase ethene hydrogenation reaction.
        }
\label{tab:c2v_ts_struct}
\begin{tabular}{  c   r@{.}l   r@{.}l  r@{.}l  }
\hline\hline
 C    &    \hspace{2mm}  -0&6834781870   &   \hspace{2mm}   0&0000000000   &   \hspace{2mm}  -0&1205362034 \\
 C    &    \hspace{2mm}   0&6834781870   &   \hspace{2mm}   0&0000000000   &   \hspace{2mm}  -0&1205362034 \\
 H    &    \hspace{2mm}  -1&5774417549   &   \hspace{2mm}   0&0000000000   &   \hspace{2mm}   1&8385459217 \\
 H    &    \hspace{2mm}   1&5774417549   &   \hspace{2mm}   0&0000000000   &   \hspace{2mm}   1&8385459217 \\
 H    &    \hspace{2mm}   1&2414310806   &   \hspace{2mm}   0&9128454494   &   \hspace{2mm}  -0&2010951235 \\
 H    &    \hspace{2mm}   1&2414310806   &   \hspace{2mm}  -0&9128454494   &   \hspace{2mm}  -0&2010951235 \\
 H    &    \hspace{2mm}  -1&2414310806   &   \hspace{2mm}  -0&9128454494   &   \hspace{2mm}  -0&2010951235 \\
 H    &    \hspace{2mm}  -1&2414310806   &   \hspace{2mm}   0&9128454494   &   \hspace{2mm}  -0&2010951235 \\
\hline\hline
\end{tabular}
\end{table}

\bibliographystyle{aip}
%\bibliography{efficient_hilbert_jagp.bib}

\end{document}